\documentclass[
    ,final            % use final for the camera ready runs
  ]
  {aipproc}
\layoutstyle{6x9}
\begin{document}
%%%%%%%%%%%%%%%%%%%%%%%%%%%%%%%%%%%%%%%%%%%%%%%%%%%%%%%%%%%%%%%%%%%%%%%%%%%%%%
\title{Underlying Event Studies for LHC Energies}


\author{Gergely G\'abor Barnaf\"oldi}{
  address={KFKI Research Institute for Particle and Nuclear Physics of the HAS, \\ 
  29-33 Konkoly-Thege M. Str. H-1121 Budapest, Hungary}
}

\author{Andr\'as G. Ag\'ocs}{
  address={KFKI Research Institute for Particle and Nuclear Physics of the HAS, \\ 
  29-33 Konkoly-Thege M. Str. H-1121 Budapest, Hungary}
  , altaddress={E\"otv\"os University, \\ 
  1/A P\'azm\'any P\'eter S\'et\'any, H-1117 Budapest, Hungary} 
}

\author{P\'eter L\'evai}{
  address={KFKI Research Institute for Particle and Nuclear Physics of the HAS, \\ 
  29-33 Konkoly-Thege M. Str. H-1121 Budapest, Hungary}
}

%%%%%%%%%%%%%%%%%%%%%%%%%%%%%%%%%%%%%%%%%%%%%%%%%%%%%%%%%%%%%%%%%%%%%%%%%%%%%%
\begin{abstract}

Underlying event was originally defined by the CDF collaboration 
decades ago. Here we improve the original definition to extend our 
analysis for events with multiple-jets.  
We introduce a definition for surrounding rings/belts and based on this definition
the jet- and surrounding-belt-excluded areas will provide a good 
underlying event definition. We inverstigate our definition via the multiplicity
in the defined geometry. In parallel, mean transverse momenta of these areas also
studied in proton-proton collisions at $\sqrt{s}=7$ TeV LHC energy. 
\end{abstract}

\maketitle
%%%%%%%%%%%%%%%%%%%%%%%%%%%%%%%%%%%%%%%%%%%%%%%%%%%%%%%%%%%%%%%%%%%%%%%%%%%%%%
\section{Introduction}

Underlying event (UE) was originally defined by the CDF 
Collaboration~\cite{CDFUE} and used to investigate properties of the remaining
of the events, after jets were identified and removed from there. The CDF
definition of the underlying event is a simple tool in order to work, however
detailed structure or information on off-jet particles cannot be obtained. On
the other hand the definition is not capable to analyze more than
2-jet structures. This motivate us to develop a new definition for the
underlying event.

To enhance the information content to be extracted from underlying events, we
modified the above CDF's definition introducing multiple surrounding belts
(SB) around the identified jets~\cite{Agocs:2009,Agocs:2010}. This new definition is
immediately leads a more detailed analysis of the underlying event, even in
case of multiple jets. On the other hand, as a specific case of our new
method, one can get the originally extracted physical observables  
corresponding to the analysis based on the CDF-definition.

In this short contribution we present the basic properties of the two 
ways of defining underlying event. We recall the original CDF-based and our 
new definition of the underlying events, which will be compared. 
We used two physical quantities for our comparison: (i) the 
average hadron multiplicity within the defined areas and (ii) the mean 
transverse momenta {\sl versus} multiplicity in the given regions. Quantities 
were investigated for both definition in parallel.

Our analysis is based on jet production and identification in proton-proton 
collisions at $7$ TeV. We used the LHC10e14 jet-jet sample  
generated by PYTHIA6.2~\cite{pythia62} 
framework with cone-based UA1~\cite{UA1} jet finder. 

%%%%%%%%%%%%%%%%%%%%%%%%%%%%%%%%%%%%%%%%%%%%%%%%%%%%%%%%%%%%%%%%%%%%%%%%%%%%
\section{Generalized definition of the underlying event}

Any definition of underlying event should strongly depend on a jet-identification 
method applied in the analysis. There are various state-of-the-art 
development on this direction~\cite{Salam:2009,Salam:2010}, which are very promising. 
On the other hand there are still a problematics of these definitions -- the 
strong process dependence. E.g. changing from proton-proton to nucleus-nucleus
collisions need to re-tune the properties of the algorithms in order to find 
and separate jets and the baseline/background of each event~\cite{Sevil}.

The CDF-based underlying event definition corresponds to the jet identification 
in case of a one- (or two-) jet events. Near side jets easily define the 
{\sl toward} and the opposite {\sl away} regions of the event geometry. Our 
original concept was to improve the CDF-based definition on a two-folded
way: 

\begin{itemize}

\item to develope an underlying event definition which is capable to 
handle multijet events.

\item to investigate the surrounding areas around identified jets, even 
without major changes of the jet-findig 
parameters in a case of nucleus-nucleus collision.

\end{itemize}

These requirements are led us to the definition of surrounding belts on the 
basis of the event-background such as 'underlying event', which completely
satisfy our requirements above.   

In our method, we are using jet-finding algorithms also. We define jets, than 
based on the physical properties of the concentrical surrounding belts and 
the remaining particle multiplicities, a better background or baseline can 
be provided. On 
the other hand the analysis of the surrounding area around the identified 
jets, can even give feedback on the goodness of the jet finding 
parameters.

\begin{figure}[h]
  \includegraphics[width=.8\textwidth]{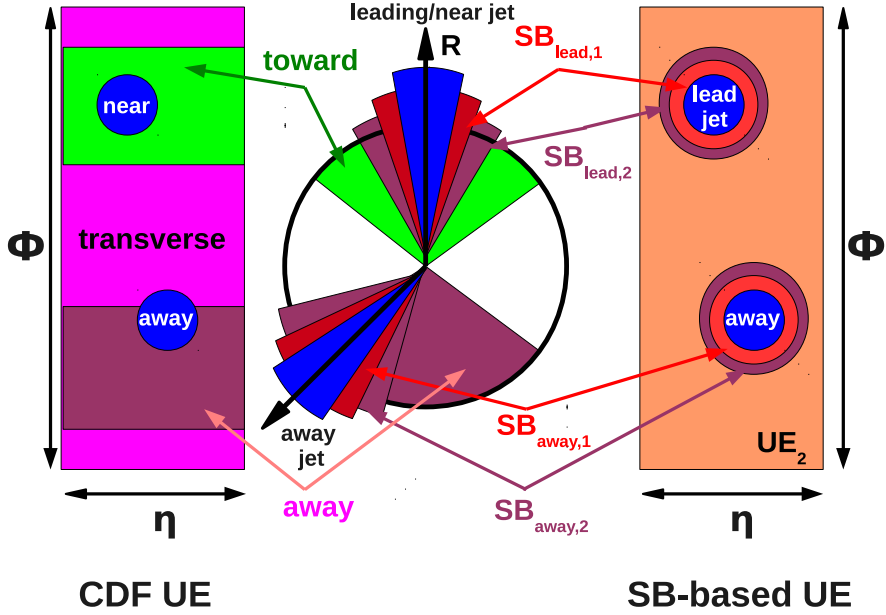}
  \caption{The schematic view of the underlying event (UE) 
defined by the CDF ({\sl left panel}) and the surrounding belts (SB, 
{\sl right panel}).  Details are in the text. (Color online.)}
  \label{fig:cdf-sb-ue}
\end{figure}

On Fig.~\ref{fig:cdf-sb-ue} the visual comparison of the two
definitions can be seen. {\sl Left side} of the figure is for the CDF-based
definition, the {\sl right side} displays 
the SB-based one. The two definitions can be summarized in a following ways, using 
the azimuth, $\Phi$ and (pseudo)rapidity, ($\eta$ or) $y$ plane: 

\begin{description}

\item[CDF-based definition] of the underlying event is based on the subtraction of 
two areas of the whole measured acceptance: one around the identified near jet 
({\sl toward region}) and another to the opposite ({\sl away}) direction. Both 
regions are $\Delta \Phi \times \Delta \eta$-slices of the measured acceptance around 
the near jet and to the opposite, with the full $\Delta \eta $ range and 
$\Delta \Phi = \pm 60^o$ in azimuth.  

\item[SB-based definition] uses all identified jets of the event to subtract 
them from the background. Each jet can have an approximate dial-like area, 
around which concentric bands (or rings) can be defined. If a 
jet cone angle, $R=\sqrt{\Delta \Phi^2 + \Delta \eta^2}$ is given, a first 
'$SB_1$' and a second '$SB_2$' surrounding belt can be defined 
for any jet with the thicknesses of $\delta R_{SB1}$ and $\delta R_{SB2}$, 
respectively. Generally, $\delta R_{SBi}=0.1$ with respect to the 
$R \approx 0.5-1$ values. It is easy to see our underlying event definition is no 
longer jet-number dependent. 

Furthermore, increasing $\delta R_{SBi}$ values, similar (but not the same) area 
can be covered as in the original CDF-based definition. In this way the two model 
can be comparable too.
\end{description}

Now we investigate the basic properties of the areas and parallel the physical 
quantities for the selected regions. 

%%%%%%%%%%%%%%%%%%%%%%%%%%%%%%%%%%%%%%%%%%%%%%%%%%%%%%%%%%%%%%%%%%%%%%%%%%%%
\section{Comparison of UE definitions}

Here we compare the details of the CDF- and surrounding belt (SB) based 
underlying event definitions. For our test we used PYTHIA6-simulated~\cite{pythia62} 
proton-proton collisions (Perugia-0 tune~\cite{Perugia0}), namely 
LHC10e14 jet-jet at 7 TeV center-of-mass energy with 
$150,000$ events. This sample contains jets identified by UA1 method~\cite{UA1}.
We restricted our analysis to the settings of $p_{THardMin} = 10$ GeV/c and 
$p_{THardMax} = 20$ GeV/c. 

\begin{figure}
  \includegraphics[width=.55\textwidth]{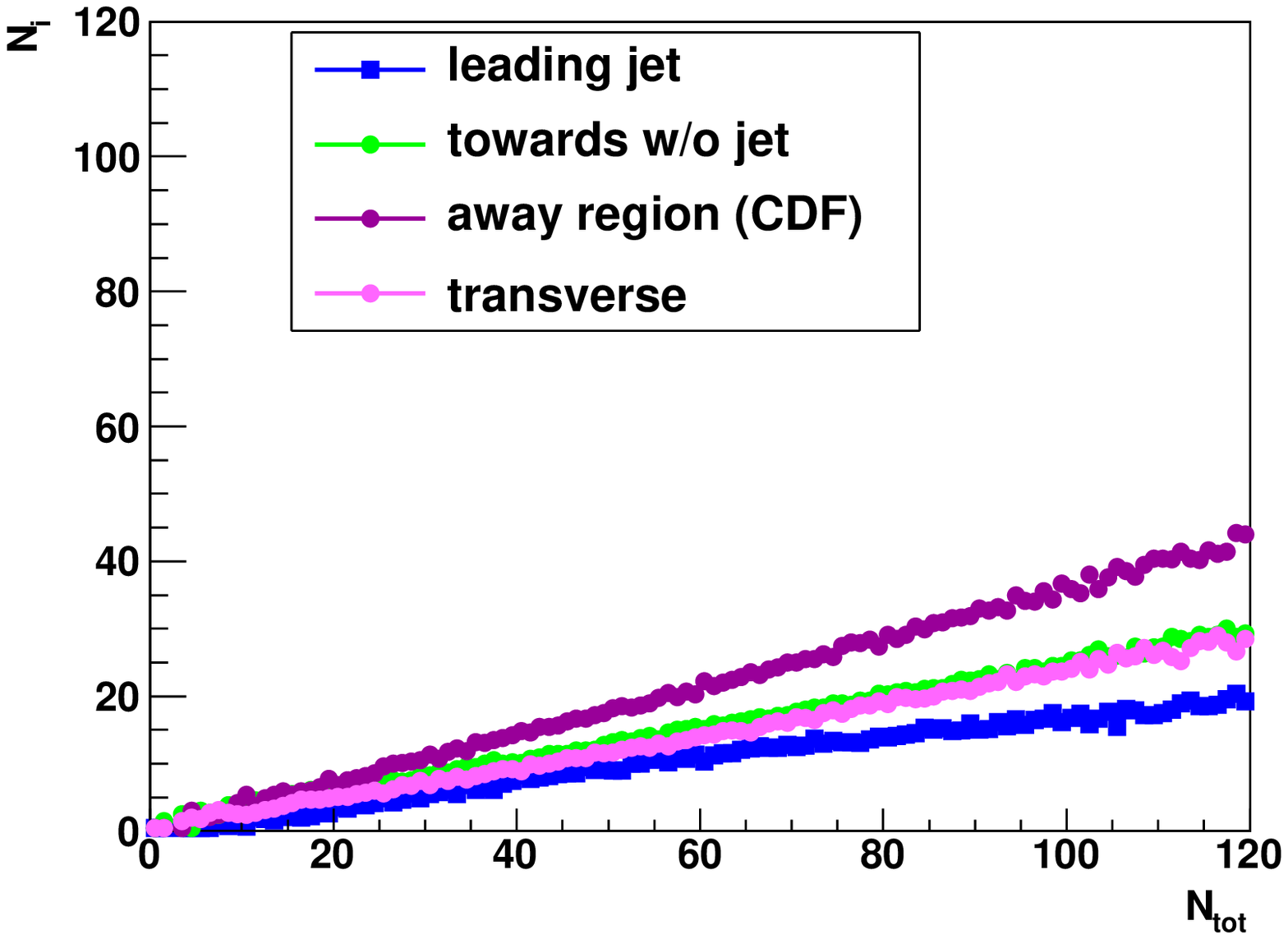}
  \hspace*{-1.3truecm}
  \includegraphics[width=.55\textwidth]{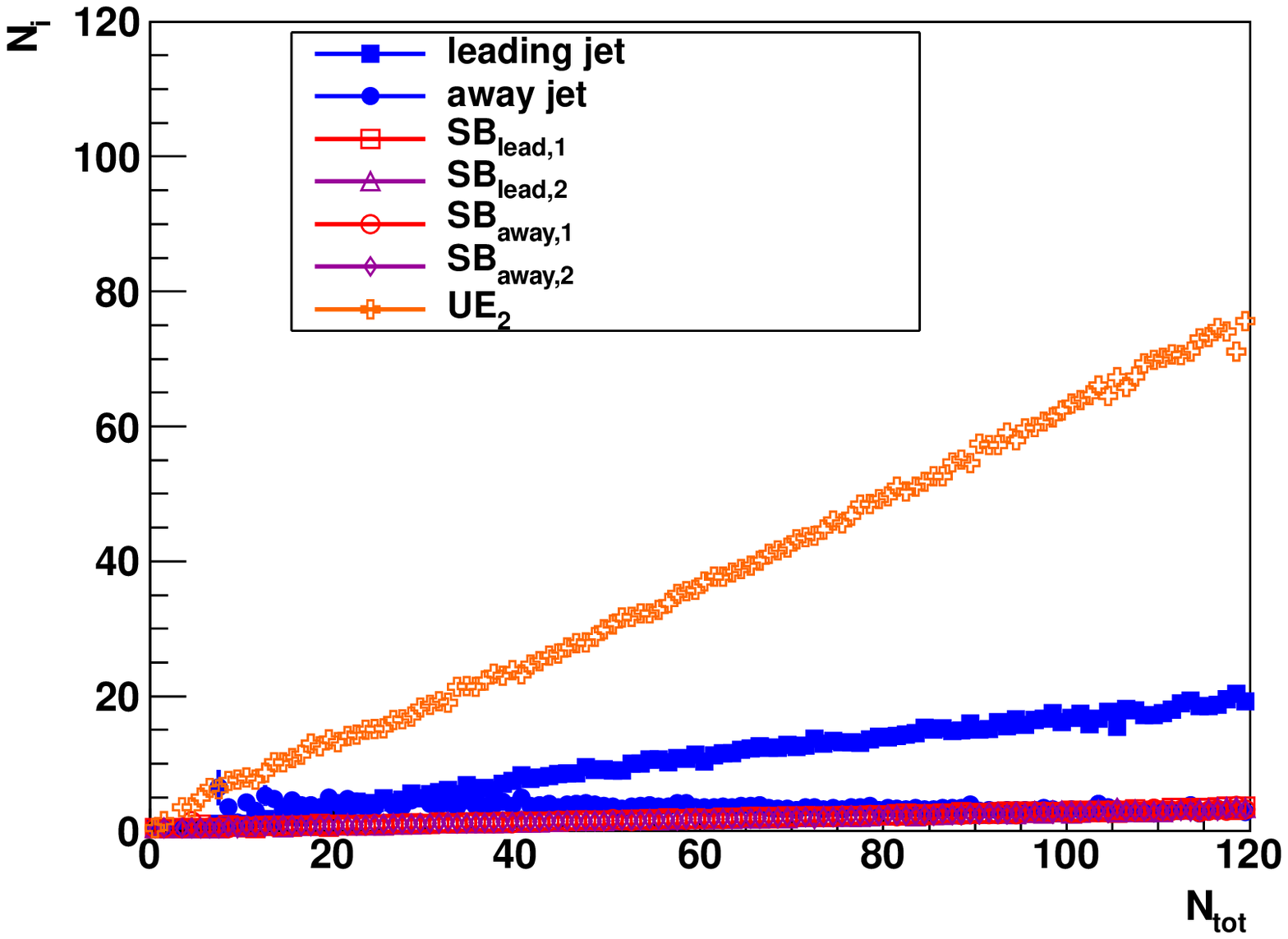}
  \caption{The multiplicity, $N_i$ for the selected areas as the 
multiplicity of the total event, $N_{tot}$. Underlying event regions are 
defined on the {\sl left panel} for the CDF-based and on the {\sl right panel} 
for the surrounding belt based definitions. More details are in the text. 
(Color online.)}
  \label{fig:n-vs-ntot}
\end{figure}

Primarily we investigated the multiplicities of various geometrical regions 
of the generated events based on the full sample. After applying UA1 
jet finding algorithm to identify jets, we compared the selected 
areas using both CDF- and SB-based definitions of the underlying events. On left
panel of Fig.~\ref{fig:n-vs-ntot} we plotted the multiplicities, $N_i$ of the 
CDF-selected areas versus the total event multiplicity. Here $N_i$ refers 
for followings: the multiplicities of the identified 'leading/near jet' 
({\sl blue squares}), 
the jet-excluded 'toward' area ({\sl green dots}), the 'away' side area to the 
opposite direction ({\sl purple dots}), and the CDF-defined underlying is
event, 'transverse' ({\sl pink dots}). 
The {\sl right side} of the Fig.~\ref{fig:n-vs-ntot} 
stands for the SB-based underlying event definition with more areas:
the multiplicities of the identified leading jet 
{\sl blue squares}), the away side jet ({\sl blue dots}), multiplicity 
for the surrounding belts, $SB_{lead,1}$, $SB_{lead,2}$, $SB_{away,1}$, 
and $SB_{away,2}$ are {\sl open red squares, open purple triangles, 
open red circles, open purple diamonds} respectively. Finally {\sl orange
crosses} denote multiplicity for the newly defined underlying event $UE_2$ 
outside all jets. (Note, all color in accordance with the areas of 
Fig~\ref{fig:cdf-sb-ue} above.) 

Fig.~\ref{fig:n-vs-ntot} shows multiplicity in almost all regions: $N_i$ increases
almost linearly with the total multiplicity, in the $N_{tot}< 120$ region 
of the event for both cases. In case of the CDF-based definition, the away region gives the 
biggest contribution, and the jet belongs to the smallest one. The transverse
(underlying event) area lies between the two extremal contribution. Moreover,
it is interesting to see, after excluding the jet from the toward region, the
remaining area has almost the same multiplicity as the underlying event. This 
shows the goodness of the jet finding algorithms and the "safety" 
of the CDF-based underlying event definition (e.g. 1/3 of the whole acceptance 
far from any jet-contaminated areas). 

The multiplicity relations of the SB-based definition differs from the
CDF-based. The near jet has the same contribution, away side jet and the 
$SB_i$s have small fraction from the $N_{tot}$ -- due to the small areas. 
On the other hand, the newly defined underlying event, $UE_2$ dominates 
the event multiplicity since it has almost the whole acceptance. 

In general the multiplicity fraction of the defined 
areas are almost proportional to the geometrical surface, only the jet-content 
part violates this dependence, as Fig.~\ref{fig:n-vs-ntot} displays. Thus, the SB-based $UE_2$ has larger 
multiplicity comparing to the CDF-based one, which might gives better statistics
for an underlying event analysis. 

\begin{figure}
  \includegraphics[width=.55\textwidth]{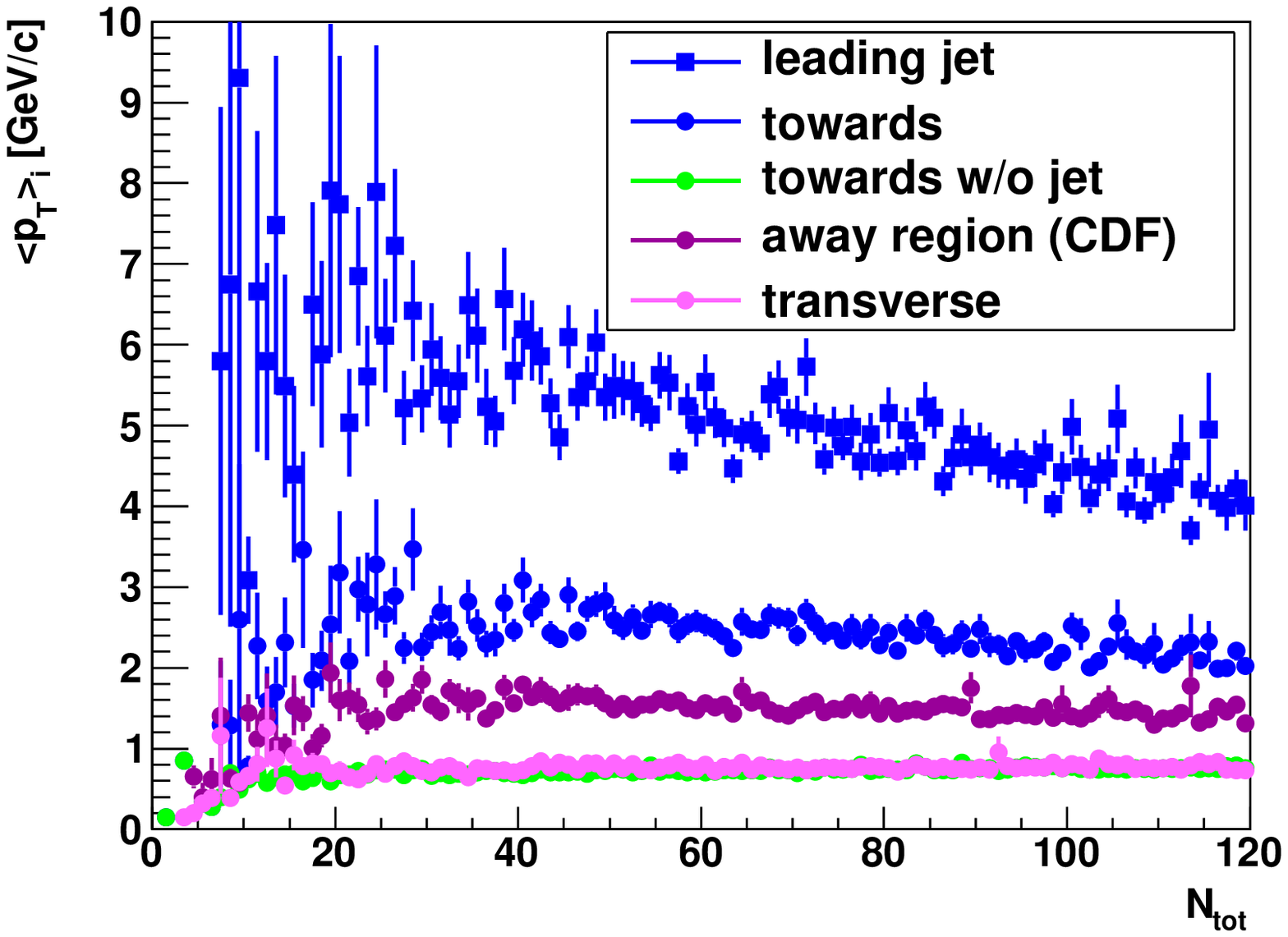}
  \hspace*{-1.3truecm}
  \includegraphics[width=.55\textwidth]{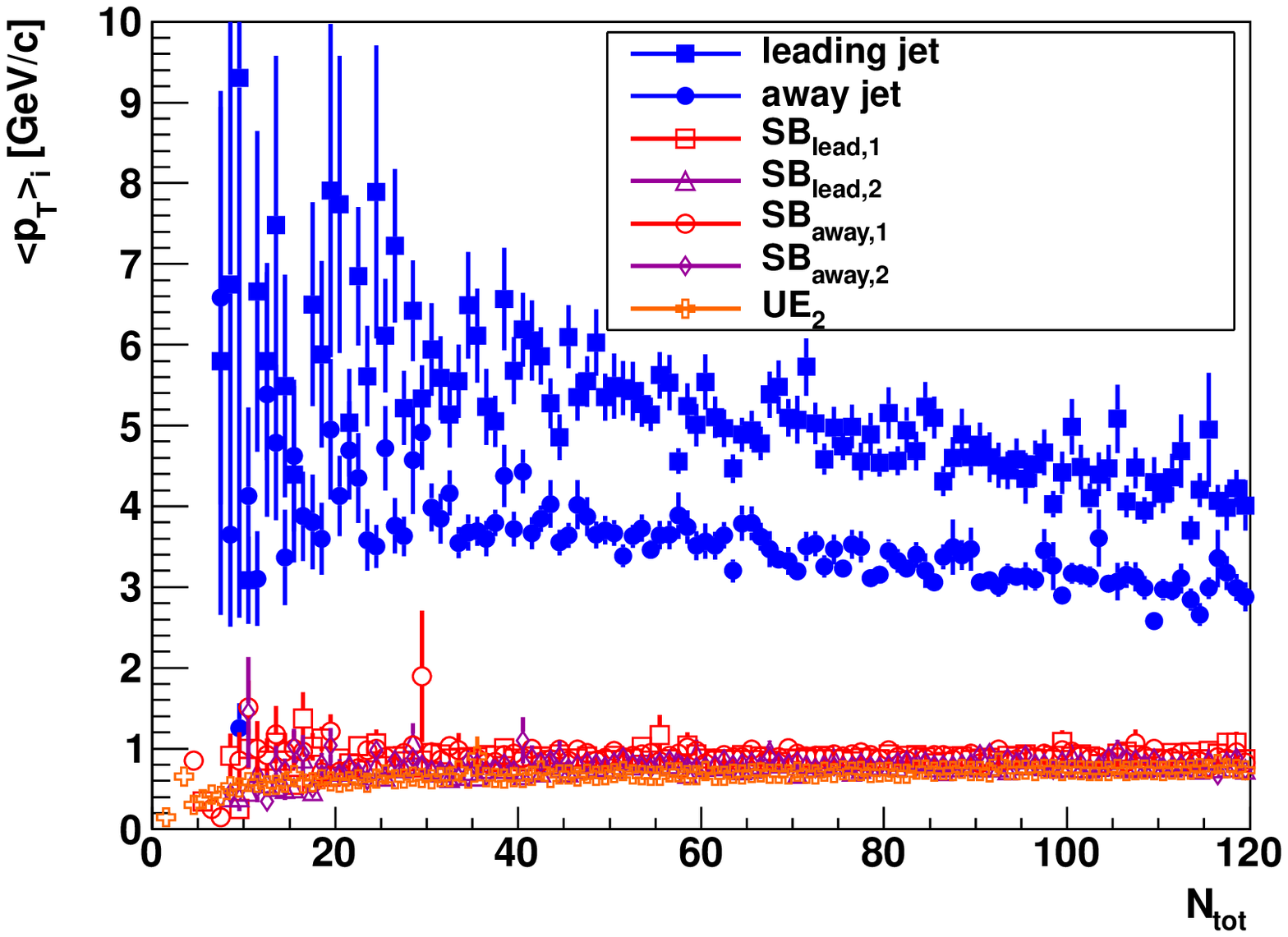}
  \caption{The average transverse momenta versus the total multiplicity of the 
events, $N_{tot}$. Regions defined on the {\sl left panel} are for the CDF-based 
and on the {\sl right panel} are for the SB-based definitions. More 
details are in the text. (Color online.)}
  \label{fig:avpt-ntot}
\end{figure}

Secondly, the mean transverse momentum $\langle p_T \rangle$ of the selected areas
is investigated including especially the underlying event.
We plotted the $\langle p_T \rangle$  vs. the multiplicity of 
the total event, $N_{tot}$ and the $\langle p_T \rangle$ vs. multiplicity of the 
CDF-based and SB-based underlying events. 

On  Fig.~\ref{fig:avpt-ntot} we display the $\langle p_T \rangle$ vs. the 
multiplicity of the total event, $N_{tot}$ for both CDF-based ({\sl left panel}) 
and SB-based ({\sl right panel}) underlying event. We use the same color and 
mark encoding for the selected regions of the event as on Fig.~\ref{fig:n-vs-ntot} 
above. 

We found the mean-$p_T$ distributions of the regions are similar in proton-proton
collisions. The identified leading jet has the highest values 
$\langle p_T \rangle_{leading \ jet} \sim  8-9$ GeV/c, which are decreasing to 
$\langle p_T \rangle_{leading \ jet} \sim 5-6$ GeV/c as going to larger $N_{tot}$, for both definition's cases.   
The mean-$p_T$ for both underlying event cases are the same with the constant 
value $\langle p_T \rangle_{UEi} \sim 0.5$ GeV/c. For the CDF-based definition this 
is similar to the jet-excluded toward area also, and for the SB-based definition
surrounding belts, $SB_i$ have also similar, but a slightly higher 
$\langle p_T \rangle_{SBi} \sim  1.0$ GeV/c. Differences between the left and the right 
panels are originating from the handling of the near and away side jet. CDF-based 
definition contains the jet to the near (leading) direction, and fully or 
partially to the opposite away region: $\langle p_T \rangle_{towards} \sim 2-3$ 
GeV/c and $\langle p_T \rangle_{away} \sim 1.5-2$ GeV/c. SB-based case since away side 
jets were also identified, $\langle p_T \rangle_{away \ jet} \sim 2-3$ GeV/c.
We can state generally, both underlying-event definition give the same result and
they are differ only in the separation (or inclusion) of leading or away 
side jet to the given areas. 

\begin{figure}[h]
  \includegraphics[width=.55\textwidth]{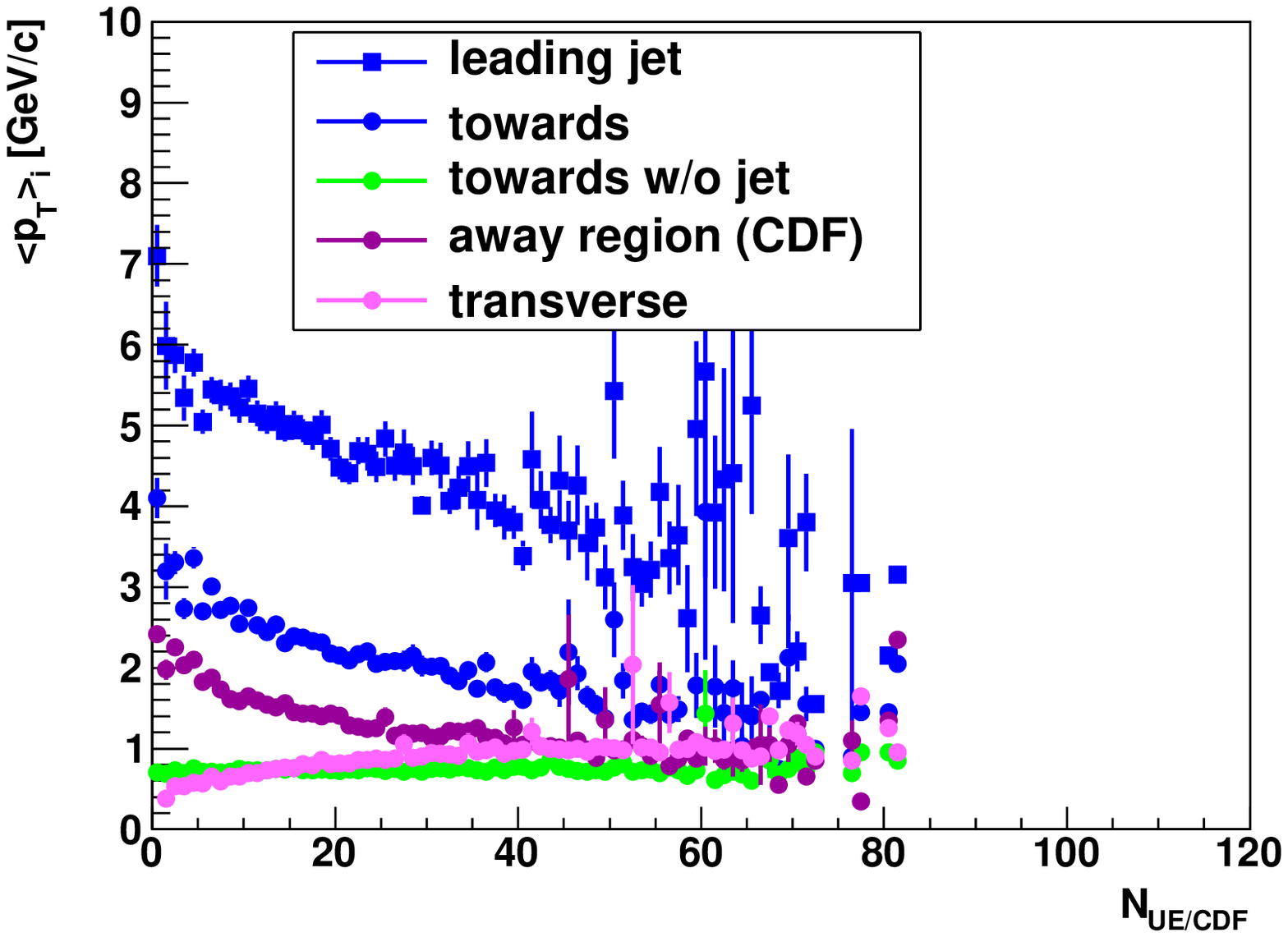}
  \hspace*{-1.3truecm}
  \includegraphics[width=.55\textwidth]{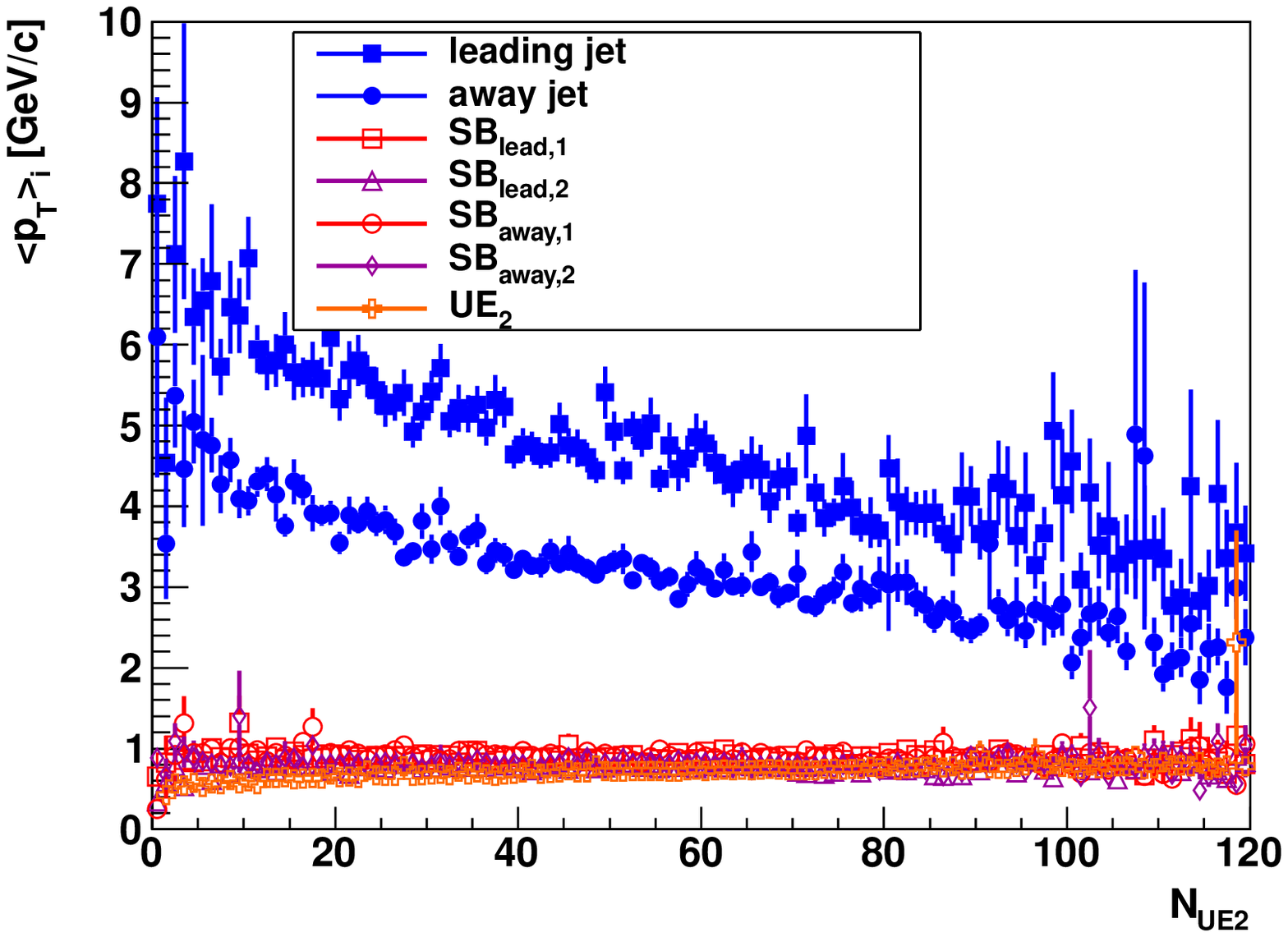}
  \caption{The average transverse momenta versus the multiplicity of the 
underlying evens. Regions are defined on the {\sl left panel} for the CDF UE 
and on the {\sl right panel} for the surrounding belt based UE definitions. More 
details are in the text. (Color online.)}
  \label{fig:avpt-nue2}
\end{figure}
Finally on Fig.~\ref{fig:avpt-nue2} we compared mean-$p_T$ values vs. to the 
self-definition-given underlying events: CDF-based UE, $N_{UE/CDF}$ on 
{\sl left panel} and SB-based $N_{UE2}$ on {\sl right panel}. (Colors and marks are
the same as above figures.) Here, the comparison shows slight difference between 
the panels. A stronger decrease in the highest $\langle p_T \rangle$-content 
regions present compared to Fig.~\ref{fig:avpt-ntot}. Furthermore, 
changing from $N_{tot}$ to $N_{UE/CDF}$ and $N_{UE2}$ the separation of the curves 
are more clear in both cases, especially at the largest $\langle p_T \rangle$ values. 
In parallel the mean-$p_T$ values for the underlying event are almost the same 
for the average multiplicity events and slightly higher for the rare ones.

%%%%%%%%%%%%%%%%%%%%%%%%%%%%%%%%%%%%%%%%%%%%%%%%%%%%%%%%%%%%%%%%%%%%%%%%%%%%
\section{Conclusions}

We studied our new underlying event definition in 
$\sqrt{s} = 7$ TeV proton-proton collisions with 150,000 events. We investigated 
and compared the multiplicities and the mean-$p_T$ vs. multiplicities for 
the CDF-based and our SB-based definition.

We found the multiplicity fraction of the defined regions are almost proportional 
to the geometrical surface, only the jet-content part differs, due to the separation
(or inclusion) of the leading or away side jets. The SB-based underlying event, 
$UE_2$ found to have larger multiplicity comparing to the CDF-based, $N_{UE/CDF}$ 
one, which might gives better statistics for the underlying event analysis. 

The mean-$p_T$ vs. $N_{tot}$ analysis led us to compare both definition on 
the same level. We got the same dependence of the underlying event for both, CDF- 
and SB-based cases. On the other hand, the above mentioned jet and near/away-area 
handling leads to differences. 

Finally, we compared our definitions by the mean-$p_T$ vs. the self-defined 
underlying event multiplicities, namely $N_{UE/CDF}$ and $N_{UE2}$. Our results have 
shown both definition is reliable, and -- due to the generalized definition of the 
surrounding belt based underlying event -- multiple jets and detailed analysis of
the surrounding areas can be performed in the future.    

%%%%%%%%%%%%%%%%%%%%%%%%%%%%%%%%%%%%%%%%%%%%%%%%%%%%%%%%%%%%%%%%%%%%%%%%%%%%
\begin{theacknowledgments}
This work was supported by Hungarian OTKA NK77816, PD73596 and
E\"otv\"os University. Authors (GGB \& PL) are appreciate the  
local support by the UNAM, Mexico and GGB thanks for the J\'anos Bolyai Research 
Scholarship of the HAS.
\end{theacknowledgments}

%%%%%%%%%%%%%%%%%%%%%%%%%%%%%%%%%%%%%%%%%%%
%% The following lines show an example how to produce a bibliography
%% without the help of the BibTeX program. This could be used instead
%% of the above.
%%%%%%%%%%%%%%%%%%%%%%%%%%%%%%%%%%%%%%%%%%%

%%%%%%%%%%%%%%%%%%%%%%%%%%%%%%%%%%%%%%%%%%%%%%%%%%%%%%%%%%%%%%%%%%%%%%%%%%%%

%%%%%%%%%%%%%%%%%%%%%%%%%%%%%%%%%%%%%%%%%%%%%%%%%%%%%%%%%%%%%%%%%%%%%%%%%%%%
%%
%% End of file `template-6s.tex'.
%%%%%%%%%%%%%%%%%%%%%%%%%%%%%%%%%%%%%%%%%%%%%%%%%%%%%%%%%%%%%%%%%%%%%%%%%%%%
\end{document}